# UNIVERSE WITHOUT SINGULARITIES
# A GROUP APPROACH TO DE SITTER COSMOLOGY

Ignazio Licata
*Isem, Institute for Scientific Methodology, Pa, Italy*
*IxtuCyber for Complex Systems*,
via Favorita 9, 91025, Marsala (TP), Italy
Ignazio.licata@ejtp.org

**Abstract:** In the last years the traditional scenario of "Big Bang" has been deeply modified by the study of the quantum features of the Universe evolution, proposing again the problem of using "local" physical laws on cosmic scale, with particular regard to the cosmological constant role. The "group extention" method shows that the De Sitter group univocally generalizes the Poincarè group, formally justifies the cosmological constant use and suggests a new interpretation for Hartle-Hawking boundary conditions in Quantum Cosmology.

## 1. Introduction

There are strong theoretical coherence reasons which impose to critically reconsider the approach to cosmological problem on the whole. The Quantum Cosmology's main problem is to individuate the proper boundary conditions for the Universe's wave function in the Wheeler-DeWitt equation. These conditions have to be such to allow the confrontation between a probability distribution of states and the observed Universe. In particular, it is expected to select a path in the configuration space able to solve the still open problems of the Big-Bang traditional scenario: flat space, global homogeneity (horizon problem) and the "ruggedness" necessary to explain the tiny initial dishomogeneities which have led to the formation of the galactic structures.

The inflationary cosmology ideas has partly supplied with a solution to the standard model wants by introducing the symmetry breaking and phase transition notions which are at the core of Quantum Cosmology. The last one also finds its motivation in the necessity to provide with a satisfactory physical meaning to the initial singularity problem, unavoidable in GR under the condition of the Hawking-Penrose theorem (Hawking & Ellis, 1973).

The Hartle-Hawking "no-boundary" condition seems to provide a very powerful constraint for the Quantum Cosmology main requirements, but appears as an "ad hoc" solution which could be deduced by a fundamental approach. Particularly, the mix of topologies used to conciliate the without boundary Universe symmetry with the Big-Bang evolutionary scenario is unsatisfactory.

We realize that most part of the Quantum Cosmology problems inherit the uncertainties of the Fridman model in GR, so they derive from the euristic use of the local laws on cosmic scale.

A possible way-out is the Fantappié-Arcidiacono group approach which allows to individuate a Universe model without recourse to arbitrary extrapolations of the symmetry groups valid in physics.

The group extension theory naturally finds again the Hartle-Hawking condition on the Universe wave function and allows to firmly founding theoretically the Quantum Cosmology. The price to pay is a subtle methodological question on using the GR in cosmology. In fact, in 1952 Fantappié pointed out that the problem of the use of *local laws* to define the cosmological boundary conditions is due to the fact that GR describes matter in terms of local curvature, but leaves the question of space-time global structure indeterminate. It happens because, differently from RR, GR has not be built on group base, which thing should be central in building any theory up, especially when it aims to express universally valid statements on physical world, the class of *the superb theories,* how Roger Penrose called them.

We are going to examine here the foundations of the group extension method (par. 2) and the relativity in the De Sitter Universe (par. 3, 4), we introduce the conditions to define matter-fields (par.5).In (par.6) we analyze the physical significance of the observers in an istantonic Universe at imaginary time, and in (par.7) investigate the physical meaning of an Hartle-Hawking condition in an hyper-spherical universe.

## 2. An Erlangen Program for Cosmology

In 1872 Felix Klein (1849-1925) presented the so-called *Erlangen program* for geometry, centred upon the symmetry transformations group. From 1952, Fantappié, basing on a similar idea and in perfect consonance with Relativity spirit, proposed an *Erlangen program for physics*, where a Universe is univocally individuated by a symmetry group which let its physical laws invariant (Fantappié,1954, 1959). It has to be underlined that in the theory *Universe* means any physical system characterized by a symmetry group.

The space-time isotropy and homogeneity principle with respect to physical laws tells us that *the physical law concept itself is based upon symmetry*. So the essential idea is to individuate physical laws starting from the transformations group which let them invariant. We observe here that there are infinite possible transformations group which individuate an isotropic and homogeneous space-time. In order to build the next improvements in physics using the *group extension* method, we can follow the path indicated by the two groups we know to be two valid description levels of the physical world: the Galilei group and the Lorentz-Poincarè one. It is useful to remember that the Galilei group is a particular case of the Lorentz one when $c \to \infty$, i.e. when it is not made use of the field notion and the interactions velocity is considered to be infinite. Staying within a quadrimensional space-time and consequently considering only groups at 10 parameters and continuous transformations, Fantappié showed that the Poincaré group can be considered a limit case of a broader group depending with continuity on $c$ and another parameter $r$: the *Fantappié group*; moreover this group cannot be further extended under the condition to stay within a group at 10 parameters.

So we have the sequence:

$$G_{1+3}^{10} \to L_{1+3}^{10} \to F_{1+3}^{10}$$

Where G is the Galilei group, L the Lorentz one and F the Fantappié *final one*, from which with $R \to \infty$, we get the L group. It is shown that such sequence of universes is *univocal*.

The Lorentz group can be mathematically interpreted as the group of roto-translations such to let that particular *object* that is the Minkowski space-time invariant. Similarly, the Fantappié group is the one of the pentadimensional rotations of a new space-time: the hyper-spherical and at constant curvature De Sitter universe (maximally symmetric). We point out we have obtained the De Sitter model without referring to the gravitational interaction, differently from the GR where the De Sitter universe is one of the possible solutions of the Einstein equations with cosmological constant. From a formal viewpoint we make recourse to pentadimensional rotations because in the De Sitter universe there *appears* a new constant $r$, which can be interpreted as the Universe radius.

The group extension *mechanism* individuates an univocal sequence of symmetry groups; for each symmetry group we have a corresponding level of physical world description and a new universal constant, so providing the most general boundary conditions and constraining the form of the possibile physical laws. The Fantappié group fixes the $c$ and $r$ constants and defines a new relativity for the inertial observers in De Sitter Universe. In this sense, the *Theory of Universes-* based on group extension method- is actually a version of what is sought for in the Holographic Principle: the possibility to describe laws and boundaries in a compact and unitary way.

In 1956 G. Arcidiacono proposed to study the De Sitter $S_4$ *absolute* universe by means of the tangent relative spaces where observers *localize* and *describe* the physical events by using the Beltrami-Castelnuovo $P^4$ projective representation in the *Projective Special Relativity,* PSR (Arcidiacono,1956; 1976; 1984).

We note that we pass from hyper-spherical $S_4$ to its real representation as hyperboloid by means of an inverse Wick rotation, rotating $it \to \tau$ and associating the great circles on the hyper-sphere with a family of geodesics on the hyperboloid. In this way, we get a realization of the Weyl principle for defining a Universe model, because it fixes a set of privileged observers (Ellis & Williams, 1988). So, the choice of $P^4$ Beltrami-Castelnuovo is equivalent to study a relativity in $S^4$.

### 3. The Fantappié Group Transformations

To study the De Sitter $S^4$ universe according to Beltrami-Castelnuovo representation we have to set the projectivities which let the Cayley-Klein interval invariant:

(1.3) $$x^2 + y^2 + z^2 - c^2 t^2 + r^2 = 0.$$

The (1.3) meets the time axis in the two $t = \pm t_0$ "singularities", where $t_0 = r/c$ is the time it takes light to run the Universe $r$ radius. In this case the singularities' meaning is purely geometrical, not physical, and they represent the hyperboloid rims (1.3), since the De Sitter universe is lacking in "structural" singularities. The $S^4$ invariant transformations are the 5-dimensional space rotations which lead on the $P^4$ observer's space the projectivities that let the (1.3) unchanged.

Let's introduce the five homogeneous projective coordinates (Weierstrass condition):

(2.3) $$\overline{x_a}\,\overline{x_a} = r^2, \quad \text{with} \quad a = 0,1,2,3,4.$$

The $x_i$ space-time coordinates, with $i = 1,2,3,4$ are:

(3.3) $$x_1 = x, \quad x_2 = y, \quad x_3 = z, \quad x_4 = ict.$$

The connection between the (2.3) and (3.3) is given by the relation:

(4.3) $$x_i = r\,\overline{x_i}/\overline{x_0}$$

from which, owing to (2.3), we get the inverse relation:

(5.3) $$\overline{x_0} = r/a, \quad \overline{x_i} = x_i/a,$$

where $a^2 = 1 + x_i x_i/r^2 = 1 + \alpha^2 - \gamma^2$, with $\vec{\alpha} = \vec{x}/r$ and $\gamma = t/t_0$.

The searched transformation between the two $O'$ and $O$ observers consequently has the form:

(6.3) $\bar{x}'_a = \alpha_{ab} \bar{x}_b$ with $\alpha_{ab}$ orthogonal matrix.

Limiting ourselves, just for simplicity reasons, to the $\bar{x}_0, \bar{x}_1, \bar{x}_4$ variables and following the standard method, also used in RR, we get 3 families of transformations:

    A)  the space translations along the x axis, given by the $(\bar{x}_0, \bar{x}_1)$ rotation:

(7.3) $\quad \bar{x}'_1 = \bar{x}_1 \cos\vartheta + \bar{x}_0 \sin\vartheta$

$\quad\quad\quad \bar{x}'_0 = -\bar{x}_1 \sin\vartheta + \bar{x}_0 \cos\vartheta$

$\quad\quad\quad \bar{x}'_4 = \bar{x}_4 .$

Using the (4.3) and putting $tg\,\vartheta = T/r = \alpha$, we get the space-time transformations with T parameter:

(8.3) $\quad x' = \dfrac{x+T}{1-\alpha x/r} \quad , \quad t' = \dfrac{t\sqrt{1+\alpha^2}}{1-\alpha x/r} .$

The (8.3) for $r$ indeterminate, i.e. $r \to \infty$, are reduced to the well-known space translations of the classical and relativistic cases, connected by the T parameter.

    B)  the $T_0$ parameter time translation, given by the $(\bar{x}_0, \bar{x}_4)$ rotation:

(9.3) $\quad \bar{x}'_4 = \bar{x}_4 \cos\vartheta_0 + \bar{x}_0 \sin\vartheta_0$

$\quad\quad\quad \bar{x}'_0 = -\bar{x}_4 \sin\vartheta_0 + \bar{x}_0 \cos\vartheta_0$

$\quad\quad\quad \bar{x}'_1 = \bar{x}_1 .$

Putting $tg\,\vartheta_0 = iT_0/t_0 = i\gamma$ we obtain:

(10.3) $\quad x' = \dfrac{x\sqrt{1-\gamma^2}}{1+\gamma t/t_0} \quad , \quad t' = \dfrac{t+T_0}{1+\gamma t/t_0} .$

Also the (10.3), when $r \to \infty$ are reduced to the known cases of classical and relativistic physics.

    C)  the V parameter inertial transformations, given by the $(\bar{x}_1, \bar{x}_4)$ rotation:

(11.3) $$\bar{x}_1' = \bar{x}_1 \cos\varphi_0 + \bar{x}_4 \sin\varphi_0$$

$$\bar{x}_4' = -\bar{x}_1 \sin\varphi_0 + \bar{x}_4 \cos\varphi_0$$

$$\bar{x}_0' = \bar{x}_0.$$

Putting $tg\varphi = iV/c = i\beta$, here we find again the Lorentz transformations:

(12.3) $$x' = \frac{x+Vt}{\sqrt{1-\beta^2}} \, , \ t' = \frac{t+V x/c^2}{\sqrt{1-\beta^2}}.$$

The (A), (B) and (C) transformations form the Fantappié projective group which for two variables $(x,t)$ and three parameters $(T,T_0,V)$, with T translations and V velocity along $x$, can be written:

(13.3) $$x' = \frac{ax + [\beta + (\alpha-\beta\gamma)\gamma]ct + bT}{b - (\alpha-\beta\gamma)a x/r + (\gamma-\alpha\beta)t/t_0}; \ t' = \frac{a\beta x/c + [1+(\alpha-\beta\gamma)\alpha]t + bT_0}{b - (\alpha-\beta\gamma)a x/r + (\gamma-\alpha\beta)t/t_0},$$

where we have put $a = \sqrt{1+\alpha^2 - \gamma^2}$ and $b = \sqrt{1-\beta^2 + (\alpha-\beta\gamma)^2}$, with $\alpha = x/r$, $\beta = V/c$ and $\gamma = t/t_0$.

For $r \to \infty$ we get $a = 1$ and $b = \sqrt{1-\beta^2}$, and from (13.3) we obtain the Poincaré group with three parameters $(T, T_0, V)$.

The Fantappié group can be synthesized by a very clear geometrical viewpoint, saying that the De Sitter universe at $1/r^2$ constant curvature shows an elliptic geometry in its hyper spatial global aspect (Gauss-Riemann) and an hyperbolic geometry in its space-time sections (Lobacevskij). Making the "natural" $r$ unit of this two geometries tend towards infinity we obtain the parabolic geometry of Minkowski flat space.

### 4. The Projective Relativity in De Sitter Universe

The Projective Special Relativity (PSR) widens and contextualizes the relativistic results in De Sitter geometry.Just like in any physics there exists a wll-defined connection between mechanics and geometry. Therefore the PSR makes use of the notion of observer's private space, redifining it on the basis of a constant curvature.

In PSR it is introduced a space temporal double scale which connects a $(\chi,\tau)$ point of $S_4$ with a $(x,t)$ one of $P_4$ by means the (1.3) projective invariant. Given a AB straight line and put as R and S the intersections with (1.3), the projective distance is given by the logarithm of the (ABRS) bi-ratio:

(1.4) $$AB = (t_0/2)\log(ABRS) = (t_0/2)\log(AR \cdot BS)/(BR \cdot AS).$$

From the (1.4) we obtain:

(2.4) $\chi = r\,arctg\dfrac{x}{r}$ and $\tau = \dfrac{t_0}{2}\log\dfrac{t_0+t}{t_0-t}$.

From the (2.4) second one, similar to the Milne's formula, we can see that the "formal" singularities are related to the projective description which depicts a universe with infinite space and finite time, whereas the De Sitter one is with finite space and infinite time. It is important to underline that such equivalence between an "evolutionary" model and a "stationary" one, differently from what is often stated, is purely geometrical and has nothing to do with the physical processes, but it deals with the cosmological observer definition. We will speak again about such fundamental point further.

The addition of durations' new law:

(3.4) $$d = \dfrac{d_1 + d_2}{1 + d_1 d_2 / t_0^2}$$

it is obtained by the (10.3) formulae and finds its physical meaning in the appearing of the new $t_0 = r/c$, interpretable as the "universe age" for any $P^4$ observer family.

Let us consider a uniform motion with $U$ velocity, given by $x' = Ut'$, by means of Fantappié transformations we have a uniform motion with $W$ velocity given by:

(4.4) $$W = \dfrac{(1+\alpha^2)U + (1-\gamma^2)V + \alpha\gamma c(1 - UV/c^2)}{a(1 + UV/c^2)}.$$

For the visible universe of the $O$ observer, inside the light-cone, it is valid the condition $\alpha = \pm\gamma$ and $a=1$, and the (4.4) can be simplified as:

(5.4) $$W = \dfrac{U + V \pm \alpha^2 c(1 + U/c)(1 - V/c)}{1 + UV/c^2}.$$

For $V = c$ then $W=c$, according to RR, while for $U=c$ we have:

(6.4) $$W = c \pm 2\alpha^2 c(1 - V/c)/(1 + V/c) \neq c.$$

The (6.4) expresses the possibility of observing hyper-$c$ velocity in PSR. The outcome is less strange than it can seem at first sight, because now the space-time of an observer is defined not only by the $c$ constant but also by $r$, and the light-cone is at variable aperture. In straighter physical terms it means that when we observe a far universe region of the $t_0 = r/c$ order, the cosmic objects' velocity appears to be superior to $c$ value, even if the region belongs to the light-cone of the observer's past. For $b=0$ we obtain the angular coefficients of the tangents to the (1.3) Cayley-Klein invariant starting from a P point of the Beltrami-Castelnuovo projection, which represent the two light-cone's straight lines. Differently from RR, here the light-cone's angle is not constant and depends on the P point according to the formula:

(7.4) $$tg\,\vartheta = 2a/(\alpha^2 + \gamma^2).$$

From the (7.4) derives the $C$ variation of the light velocity with time:

$$(8.4) \quad C = \frac{c}{\sqrt{1-\gamma^2}}, \text{ with } \gamma = \frac{t}{t_0} = \frac{ct}{r},$$

from which follows that $C \to \infty$ in the two $\pm t_0$ singularities which fix the limit duration according to the addition of durations' new law (3.4).

Another remarkable consequence of the projective group is the expansion-collapse law, that is the connection between the two singularities. Differentiating the (10.3) and dividing them we obtain the velocities' variation law for a translation in time:

$$(9.4) \quad V'\sqrt{1-\gamma^2} = V(1+\gamma t/t_0) - \gamma x/t_0.$$

For $\gamma = 1$ and $T_0 = t_0$ we have the law of projective expansion valid for $-t_0 < t < 0$:

$$(10.4) \quad V = \frac{x}{t+t_0}, \text{ or also } \beta = \frac{\alpha}{1+\gamma}.$$

If $\gamma = 0 (t = 0)$, we can write

$$(11.4) \quad V = x/t_0 = Hx, \ (\beta = \alpha),$$

where $H = c/r = 1/t_0$ is the well-known Hubble constant.

The analogous procedure will be followed for the law of projective collapse valid for $0 < t < t_0$, with $\gamma = -1$ and $T_0 = -t_0$:

$$(12.4) \quad V = \frac{x}{t-t_0}, \text{ or } \beta = \frac{\alpha}{\gamma - 1}.$$

We note that in singularities the expansion-collapse velocity becomes infinite. In PSR such process, differently from GR, is not connected to gravitation, but derives from Beltrami-Castelnuovo geometry.

From the Fantappié group it also follows a new formula for the Doppler effect:

$$(13.4) \quad \omega' = \omega\sqrt{(1-\beta)/(1+\beta) + \alpha^2},$$

where $\omega$ is the frequency. For $\beta = 1$, which is V=c, we get nothing but the traditional proportionality between distance and frequency, $\omega' = \alpha\omega$. For V=0 there follows a Doppler effect depending on distance:

$$(14.4) \quad \omega' = \omega\sqrt{1+\alpha^2}.$$

The $z$ red-shift is defined by $1+z = \omega/\omega'$ and the (13.4) becomes:

$$(15.4) \quad 1/(1+z) = \sqrt{(1-\beta)/(1+\beta) + \alpha^2},$$

which was historically introduced- in a 1930 Accademia dei Lincei famous memoir- by Castelnuovo to explain the "new" Hubble observations on galactic red-shift. If we are placed on the observer's light-cone where the (12.4) becomes $\beta = \alpha/(1-\alpha)$, the (15.4) will be:

(16.4) $\qquad 1 + z = 1/(1-\alpha)$.

The red-shift tends towards infinity for $x = r$, and hyper-$c$ velocities are possible if $z > 1$.

As everybody would naturally expect, modifying geometry implies, as well as in RR, a deep redefinition of mechanics. In PSR, the $m$ mass of a body varies with velocity and distance according to:

(17.4) $\qquad m = m_0\, a^2/b$.

From the (17.4) it follows that for $a = 0$, in singularities, the mass is null, while on the light-cone, for $b = 0$, $m \to \infty$. The mass of a body at rest varies with $t$ according to:

(18.4) $\qquad m = m_0\left(1 - \gamma^2\right)$,

from which we deduce that at the initial and final instant, $\gamma = \pm 1$, the mass vanishes.

Another greatly important outcome (Arcidiacono, 1977) is the relation between $m$ mass and the $J$ polar inertia momentum of a body:

(19.4) $\qquad J = mr^2$

A remarkable consequence is that the universe $M$ mass varies with $t$:

(20.4) $\qquad M(t) = M_0\left(1 - \gamma^2\right) + \dfrac{J}{r^2}$,

where $M_0$ is the mass for $t = 0$, and $J$ the polar momentum with respect to the observer.

So the overall picture for an inertial observer in a De Sitter Universe is that of a universe coming into existence in a singularity at $-t_0$ time, expanding and collapsing at $t_0$ time and where $c$ light velocity is only locally constant. In the initial and final instants the light velocity is infinite and the global mass is zero while in the expansion-collapse time it varies according to (20.4). In the projective scenario the space flatness is linked to the observer geometry in a universe at constant curvature. All this is linked to the fact that in PSR the translations and rotations are indivisible. In the singularities there is no "breakdown" of the physical laws because the global space-time structure is univocally individuated by the group which is independent of the matter-energy distribution. In this case, the singularities in $P^4$ are – more properly- an horizon of events with a natural "cosmic censure" fixed by observers' geometry.

## 5. The Projective Gravitation

The connection between the metric approach to Einstein gravitation and Fantappié-Arcidiacono group one is the aim of Projective General Relativity(PGR), which describes a universe globally at constant curvature and locally at variable curvature. It can be done by following the Cartan idea, where any $V^4$ Riemann manifold is associated with an infinite family of Euclidean, pseudo-Euclidean, non-Euclidean spaces tangent to it in each of its P points. Those spaces' geometry is individuated by a holonomy group. The Cartan connection law links the tangent spaces so as to obtain both the $V^4$ local characteristics (curvature and torsion) and the global ones (holonomy group). The GR holonomy group is the one at four dimension rotations, i.e. the Lorentz group. So we get a general method which builds a bridgeway up between differential geometry and group theory (Pessa, 1973; Arcidiacono, 1986)

To make a PGR it is introduced the $V^5$ Riemann manifold which allows as holonomy group the De Sitter-Fantappié one, isomorphic to the $S^5$ five-dimensional rotations' group. The $V^5$ geometry is successively written in terms of Beltrami projective inducted metric for a anholomonous $V^4$ manifold at variable curvature. The Veblen projective connection:

$$(1.5) \qquad \pi^A_{BC} = \{^A_{BC}\} = \frac{1}{2} g^{AS} \left( \bar{\partial}_C g_{BS} + \bar{\partial}_B g_{CS} - \bar{\partial}_S g_{BC} \right)$$

defines a projective translation law which let the field of the $Q$ quadrics invariant in the tangent spaces, in each $V^4$ point, $Q = g_{AB} \bar{x}^A \bar{x}^B = 0$, where $g_{AB}$ are the coefficients of the five-dimensional metric, the $\bar{x}^K$ are the homogeneous projective coordinates, and $(ABC)=0,1,..,4$. From the (1.5) we build the projective torsion-curvature tensor:

$$(2.5) \qquad R^A_{BCD} = \bar{\partial}_C \pi^A_{BD} - \bar{\partial}_D \pi^A_{BC} + \pi^A_{SC} \pi^S_{BD} - \pi^A_{SD} \pi^S_{BC}.$$

So the gravitation equations of Projective General Relativity are:

$$(3.5) \qquad R_{AB} - \frac{1}{2} R g_{AB} = \chi T_{AB},$$

with $T_{AB}$ energy-momentum tensor, and $\chi$ Einstein gravitational constant. The (2.5) tensor is projectively flat, i.e. when it vanishes we get the De Sitter space at constant curvature. The deep link between rotations and translations in $S^4$ naturally leads the (3.5) to include the torsion, showing an interesting formal analogy with Einstein-Cartan-Sciama-Kibble spin-fluids theory. The construction is analogous to the GR one, but in lieu of the relation between Riemann curvature and Minkowski s-t, we get here a curvature-torsion connected to the De Sitter-Fantappié holonomy group. It has to be noted that, in concordance with the equivalence principle, the PGR gives a metric description of the *local* gravity, valid for single (i.e., non cosmological) systems.

It is here proposed again the problem of the relations between local physics and its extension on cosmic scale. In fact, if we take the starting expression of standard cosmology based upon GR, i.e. *let us consider the whole matter of Universe*, and transfer it within the ambit of PGR, we can ask ourselves if the torsion role, associated to the rotation one, could get a feed-back on the background metric, modifying it deeply. Generally, the syntax of a purely group-based theory does not get the tools to give an answer, because it is independent from gravity and the hypotheses on $T_{AB}$. For example, Snyder (Snyder, 1947) showed that in a De Sitter space it is introduced an uncertainty relation linked to a curvature of the kind: $\Delta x^i \Delta x^k \approx 1/r^2$. Only a third quantization formalism, able to take into account the dynamical two-way inter-relations between local and global, will succeed in giving an answer.

The essential point we have to underline here is that the introduction of a cosmological constant, both as additional hypothesis on Einstein equations or via group, is a radical alternative to the "machian philosophy" of the GR.

So, for a Universe without metter-fields we assume the constant curvature as a sort of "pre-matter" which describes in topological terms the most general conditions for the quantum vacuum. Therefore the Einstein equations in the following form are valid:

(4.5) $\quad G^{AB} = \Lambda g^{AB}$ and $R^{AB} = (R/2 - \Lambda)g^{AB}$,

with their essentially physical content, i.e. the deep connection among curvature, radius and matter-energy's density $\rho_{vac}$ by means of the cosmological constant:

(5.5) $\quad \rho_{vac} = \dfrac{c^2 \Lambda}{8\pi G}$.

## 6. De Sitter Observers, Singularities and Wick Rotations

From a quantum viewpoint the $S^4$ interesting aspect is that it is at imaginary cyclic time and without singularities. It means that it is impossible to define on De Sitter a global temporal coordinate. So it has an istanton feature, individuated by its Euler topological number which is 2 (Rajaraman,1982). This leads to a series of formal analogies both with black holes' quantum physics and the theoretical proposals for the "cure" for singularities.

Let us consider the De Sitter-Castelnuovo metric in real time:

(1.6) $\quad ds^2 = -\left(1 - \dfrac{H^2}{c^2}r^2\right)dt^2 + \left(1 - \dfrac{H^2}{c^2}r^2\right)^{-1} dr^2 + r^2 d\Omega^2$,

where $d\Omega^2 = d\vartheta^2 + \sin^2\vartheta d\varphi^2$ in polar coordinates.

As we have seen in PSR, the singularity in $r = c/H$ becomes an horizon of events for any observer when it passes to the Euclidean metric with $\tau \to -it$:

(2.6) $\quad ds^2 = d\tau^2 + \dfrac{1}{H^2}\cos H\tau \left(dr^2 + \sin^2 r d\Omega^2\right)$,

with a close analogy with the Schwarzschild solution's case. The $\tau$ period is $\beta = 2\pi/H$; for the observers in De Sitter it implies the possibility to define a temperature, an entropy and an area of the horizon, respectively given by:

(3.6) $\quad T_b = \dfrac{H}{2\pi} = \beta^{-1}; \quad \dfrac{\pi}{H^2} = \dfrac{\beta^2}{4\pi}; \quad A = \dfrac{4\pi}{H^2} = \dfrac{\beta^2}{\pi}$.

From the (3.6) we get the following fundamental outcome:

(4.6) $\quad S = \dfrac{1}{4} A$,

which is the well-known expression of the t'Hooft-Susskind-Bekenstein Holographic Principle(Susskind,1995). The (4.6) connects the non-existence of a global temporal coordinate with the information accessible to any observer in the De Sitter model. In this way we obtain a deep physical explanation for applying the Weyl Principle in the De Sitter Universe, and sum up that in cosmology, as well as in QM, a physical system cannot be fully specified without defining an observer. G. Arcidiacono stated that the hyper-spherical Universe is like *a book written with seven seals* ( Apocalypse, 6-11), and consequently two operations are necessary to investigate its physics: 1) inverse Wick rotation and 2) Beltrami-Castelnuovo representation. That's the way we can completely define a relativity in De Sitter.

The association of imaginary time with temperature gets a remarkable physical significance which implies some considerations on the statistical partition function (Hawking, 1975). For our aims it will be sufficient to say that such temperature is linked to the (4.6) relation, i.e. to the information that an observer spent within his area of events. Which thing has patent implications from the dynamical viewpoint, because it is the same as to state that, as well as in Schwarzschild black hole' s case, the De Sitter space and the quantum field defined on it behave as if they were immersed in background fluctuations. The transition amplitude from a configuration of a $\phi$ generic field in $t_2 - t_1 = dt$ time will be given by the $e^{-iHdt}$ matrix element which acts as a $U(1)$ group transformation of the $U(1)_{space} \Leftrightarrow U(1)_{time}$. It means that a transition amplitude on $S^4$ will appear to an observer as the $R(t)$ scale factor's variation with $H$ variation rate.

It makes possible to link the hyper-spherical description with the Big-Bang evolutionary scenario and to get rid of the thermodinamic ambiguities which characterize its "beginning" and "ending" notions. The last ones have to be re-interpretated as purely quantum dynamics of the matter-fields on the hyper-sphere free of singularities.

## 7. Physical Considerations for Further Developments

Such considerations suggest a research program we are going here to shortly delineate ; it furthermore develops the analogy between black holes, istantons and De Sitter Universes (see – for example – Frolov, Markov, Mukhanov, 1989;Strominger, 1992). It is known that the Hartle-Hawking proposal of "no-boundary" condition removes the initial singularity and allows to calculate the Universe wave function (Hartle-Hawking, 1989). In fact, it is possible – as in the usual QFT- to calculate the path integrals by using a Wick rotation as "Euclidization" procedure. In such way also the essential characteristics of the inflationary hypotheses are englobed (A. Borde, A. Guth and A. Vilenkin, 2003). The derived formalism is similar to that used in the ordinary QM for the tunnel effect, an analogy which should explain the physics at its bottom (Vilenkin, 1982; S.W. Hawking and I.G. Moss, 1982).

The group extension method provides this procedure with a solid foundation, because the De Sitter space, maximally symmetric and simply connected, is univocally individuated by the group structure, and consequently is directly linked to the space-time homogeneity and isotropy principle with respect to physical laws. The original Hartle-Hawking formulation operates a mix of topologies hardly justified both on the formal level and the conceptual one. The "no-boundary" condition is only valid if we works with imaginary time, and the theory does not contain a strict logical procedure to explain the passage to real time. This corresponds to a quite vague attempt to conciliate an hyper-spherical description at imaginary time with an evolutive one at real time according to the traditional Big-Bang scenario.In fact, it has been observed that the Hartle-Hawking condition is the same as to substitute a singularity with a "nebulosity".

The spontaneous proposal, at this point, is considering the Hartle-Hawking conditions on primordial space-time as a consequence of a global charaterization of the hyper-sphere and directly developing quantum physics on $S^4$.Which thing does not contradict the quantum mechanics formulation and its fundamental spirit, which is to say the Feynman path integrals. In other words, quantum mechanics has not to be applied to cosmology for the Universe smallness at its beginning, but because each physical system – without exception- gets quantum histories with amplitude interferences. We point out that such view is in perfect consonance with the so-called quantum mechanics Many Worlds Interpretation ( Halliwell, 1994). The "by nothing creation" means that we cannot "look inside" an istanton (hyper-spherical space), but we have to recourse to an "evolutionary" description which separates space from time. The projective methods tell us how to do it.

An analogous problem– to some extent – is that of the Weyl Tensor Hypothesis. Recently, Roger Penrose has suggested a condition on the initial singularity that, within the GR, ties entropy and gravity and makes a time arrow emerge (Penrose,1989). It is known that the $W_{ABCD}$ Weyl conformal tensor describes the freedom degrees of the gravitational field. The Penrose Hypothesis is that $W_{ABCD} \to 0$ in the Big-Bang, while $W_{ABCD} \to \infty$ in the Big-Crunch. The physical reason is that in the Universe's initial state we have an highly uniform matter distribution at low entropy
( entalpic order), while in Big-Crunch, just like a black hole, we have an high entropy situation. This differentiates the two singularities and provides a time arrow. In an hyper-spherical Universe there is no "beginning" and "ending", but only quantum transitions.Consequently, the Penrose Hypothesis can only be implemented in terms of projective representation within the ambit of PGR.

Finally, we can take into consideration the possibility to build a Quantum Field Theory on $S^4$. A QFT, for T tending towards zero, is a limit case of a theory describing some physical fields interacting with an external environment at T temperature. Without this external environment we could not speak of dechoerence , could not introduce concepts such as like dissipation, chaos, noise and, obviously, the possibility to describe phase transitions would vanish too. Therefore, it is of paramount importance to write a QFT on De Sitter background metric and then studying it in projective representation. If we admit decoherence processes on $S^4$, it is possible to interpret the Weyl Principle as a form of Anthropic Principle: the "classical" and observable Universes are the ones where it can be operated a description at real time.

In conclusion, it is possible to delineate an alternative, but not incompatible with traditional cosmology scenario.The Universe is the quantum configuration of the quantum fields on $S^4$.Thus developing a Quantum Cosmology coincides with developing a Quantum Field Theory on a space free of singularities.The Big-Bang is a by vacuum nucleation in an hyper-spherical background at imaginary time, and so the concepts of "beginning", "expansion" and "ending" belong to the space-time foreground and gain their meaning only by means of a suitable representation which defines a family of cosmological observers.

**Acknowledgements:**
I owe my knowledge of the group extension method to the regretted Prof. G. Arcidiacono (1927 – 1998), during our intense discussions while strolling throughout Rome.
Special thanks to my friends E. Pessa and L. Chiatti for the rich exchange of viewpoints and e-mails.